\documentclass[final,sort&compress]{aipproc}

\layoutstyle{8x11double}
\newcommand{\etal}{et~al{.}}

\begin{document}

\title{Nuclear spin effects in negatively charged InP quantum dots}

\classification{78.67.Hc, 71.35.Pq, 72.25.Fe, 72.25.Rb, 71.70.Jp}
\keywords{Electron Spin, Nuclear Spin, Hyperfine Interaction, InP
Quantum Dots}

\author{S. Yu. Verbin}{
  address={V. A. Fock Institute of Physics, St-Petersburg University,
  St-Petersburg 198504, Russia},
  altaddress={Venture Business Laboratory, University of Tsukuba,
  Tsukuba 305-8571, Japan}
}

\author{B. Pal}{
  address={Institute of Physics, University of Tsukuba,
  Tsukuba 305-8571, Japan}
}

\author{M. Ikezawa}{
  address={Institute of Physics, University of Tsukuba,
  Tsukuba 305-8571, Japan}
}

\author{I. V. Ignatiev}{
  address={V. A. Fock Institute of Physics, St-Petersburg University,
  St-Petersburg 198504, Russia},
  altaddress={Venture Business Laboratory, University of Tsukuba,
  Tsukuba 305-8571, Japan}
}

\author{Y. Masumoto}{
  address={Institute of Physics, University of Tsukuba,
  Tsukuba 305-8571, Japan}
}

\begin{abstract}
Effects of both the dynamic nuclear polarization (DNP) created by
circularly polarized light and the fluctuations of average nuclear
spin in a quantum dot (QD) on the electron spin orientation are
studied for singly negatively charged InP QDs. From the dependence
of the negative circular polarization of photoluminescence on the
applied longitudinal magnetic field, the hyperfine field
$B_{\mathrm{N}}$ of \emph{a few} mT appearing due to DNP and the
effective magnetic field $B_{\mathrm{f}}$ of \emph{a few tens} of
mT arising from nuclear spin fluctuations (NSF) are estimated. A
lifetime of about 1~$\mu$s is estimated for NSF.
\end{abstract}

\maketitle

Strong localization of electrons in quantum dots (QDs) may enhance
hyperfine interaction of electron spins with those of
nuclei~\cite{gammonprl86}. Various aspects of the hyperfine
interaction of electron and nuclear spins have been studied for
last three decades in different materials~\cite{optorientbook},
including InP QDs~\cite{dzhioevjetpl68}. Charge-tunable InP QDs
with one resident electron per QD, on an average, have recently
attracted considerable research interests due to the observation
of millisecond range spin lifetime of resident electrons in these
QDs~\cite{ikezawaprb72,paljpsj75}. This observation makes it a
promising candidate for quantum memory element in the emerging
fields of quantum information technology and
spintronics~\cite{spintronicsbook}. However, the influence of the
hyperfine interaction between electron and nuclear spins on the
long-lived electron spin orientation needs to be clarified.

Two effects of the electron-nuclear spin-spin interactions are
possible. One of them is the so-called dynamic nuclear
polarization (DNP). In the optical orientation of electron spins,
the spin-polarized electrons dynamically polarize the nuclear
spins due to the hyperfine coupling of the electron and nuclear
spin subsystems~\cite{optorientbook}. In turn, the spin polarized
nuclei produce an internal magnetic field $B_{\mathrm{N}}$, which
may influence electron spin dynamics. In presence of an externally
applied magnetic field $B_{\mathrm{ext}}$, electron spin subsystem
should feel an effective magnetic field $B_{\mathrm{eff}} =
B_{\mathrm{ext}} + B_{\mathrm{N}}$.

Another effect arises from the nuclear spin fluctuations (NSF).
Due to limited number of nuclear spins, typically $n \sim 10^5$,
interacting with the electron spin in a QD, random correlation of
nuclear spins may create a fluctuating nuclear polarization,
$\Delta S_{\mathrm{N}} \propto S_{\mathrm{N}}/\sqrt{n}$, where
$S_{\mathrm{N}}$ is the total spin of the polarized nuclei.
Fluctuation $\Delta S_{\mathrm{N}}$ acts on the electron spin
subsystem as another internal magnetic field, $B_{\mathrm{f}}$,
with random magnitude and orientation~\cite{merkulovprb65}.
Electron spin precession in this field results in the dephasing of
electron spins in the QD ensemble and in the three-fold decrease
in magnitude of the total electron spin
polarization~\cite{merkulovprb65,braunprl94}.

In the present paper we describe our experimental study of nuclear
spin effects on long-lived spin polarization of resident electrons
observed recently~\cite{ikezawaprb72,paljpsj75} in singly
negatively charged InP QDs.

The sample consists of a single layer of self-assembled InP QDs
embedded between GaInP barriers grown on a $n^{+}$-GaAs substrate.
The average base diameter (height) of the QDs is about 40~(5)~nm
with an areal density of about $10^{10}$~cm$^{-2}$.
Semi-transparent indium-tin-oxide electrode was deposited on top
of the sample to control the charge state of the dots by means of
applied electric bias~\cite{ikezawaprb72,paljpsj75}. For the
present study on \emph{singly negatively} charged QDs we apply an
electric bias of $U_{\mathrm{b}} = -0.1$~V, as it was found from a
previous study of trionic quantum beats~\cite{kozinprb65} on the
same sample that at this bias the QDs contain one resident
electron per dot (on an average).

\begin{figure}
\includegraphics[clip,width=6.0cm]{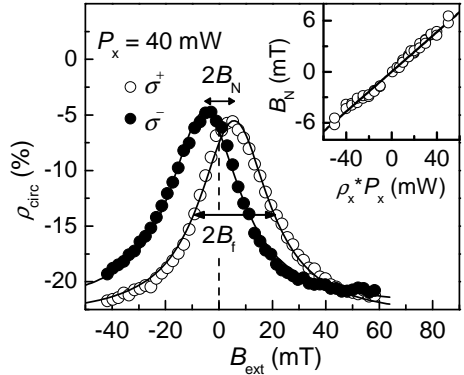}
\caption{Dependence of $\rho_{\mathrm{circ}}$ on
$B_{\mathrm{ext}}$. Inset: Excitation power dependence of
$B_{\mathrm{N}}$.} \label{fig1}
\end{figure}

Electron spins in the QD ensemble were polarized in our
experiments by using the well-known optical orientation
technique~\cite{optorientbook,ikezawaprb72,paljpsj75}. A
\emph{negative circular polarization} of the trionic (negatively
charged exciton) photoluminescence (PL) with absolute value up to
\emph{a few tens} of percentage was observed for such QDs under
quasi-resonant excitation and was interpreted as a result of
long-lived (hundreds of microseconds) spin memory of resident
electrons~\cite{ikezawaprb72,paljpsj75}. We monitor the degree of
circular polarization
$\rho_{\mathrm{circ}}=(I_{+}^{+}-I_{+}^{-})/(I_{+}^{+}+I_{+}^{-})$,
where $I_{+}^{+(-)}$ is the PL intensity for $\sigma^{+}$
excitation and detection of $\sigma^{+(-)}$ PL, as a function of
an external magnetic field $B_{\mathrm{ext}}$ applied along the
optical excitation axis (longitudinal magnetic field, Faraday
geometry). The internal magnetic field $B_{\mathrm{f}}$ is
expected to cause a three-fold reduction in the electron spin
polarization (and correspondingly in $\rho_{\mathrm{circ}}$) when
$B_{\mathrm{eff}} = B_{\mathrm{ext}} + B_{\mathrm{N}} \approx 0$.
Influence of $B_{\mathrm{f}}$ may be suppressed by applying a
longitudinal magnetic field exceeding
$B_{\mathrm{f}}$~\cite{merkulovprb65,braunprl94}. A clear
demonstration of this effect is seen in Fig.~\ref{fig1}, where
$\rho_{\mathrm{circ}}$ is plotted as a function of
$B_{\mathrm{ext}}$ for $\sigma^{+}$ and $\sigma^{-}$ excitations.
For $B_{\mathrm{ext}} > 50$~mT, $\rho_{\mathrm{circ}}$ approaches
a constant value and a decrease from this constant value is seen
for $B_{\mathrm{ext}}$ nearly, but not exactly zero. The behavior
of $\rho_{\mathrm{circ}}$ as a function of $B_{\mathrm{ext}}$ is
fitted with a Lorentzian, full width at half maximum of which
estimates $B_{\mathrm{f}} \approx 15$~mT. The shift of the
Lorentzian from $B_{\mathrm{ext}}=0$ is due to $B_{\mathrm{N}}$.
The sign of $B_{\mathrm{N}}$ created by light should be opposite
for the two counter circular polarizations ($\sigma^{+}$ and
$\sigma^{-}$) of the excitation beam. As a result, the values of
$B_{\mathrm{ext}}$ corresponding to $B_{\mathrm{eff}} =
B_{\mathrm{ext}} + B_{\mathrm{N}} \approx 0$ for $\sigma^{+}$ and
$\sigma^{-}$ excitations differ by $2B_{\mathrm{N}}$
[Fig.~\ref{fig1}]. A plot of $B_{\mathrm{N}}$ as a function of the
product $\rho_{\mathrm{x}} * P_{\mathrm{x}}$ [$\rho_{\mathrm{x}} =
+1$~($-1$) for $\sigma^{+}$~($\sigma^{-}$) excitation and
$P_{\mathrm{x}}$ is the excitation power] shows that DNP builds up
linearly with the excitation laser power [Fig.~\ref{fig1}(inset)].
Even at high excitation power of about 50~mW a rather small
dynamic nuclear magnetic field of about 6~mT is seen, in agreement
with a previous report~\cite{dzhioevjetpl68}.

\begin{figure}
\includegraphics[clip,width=5.8cm]{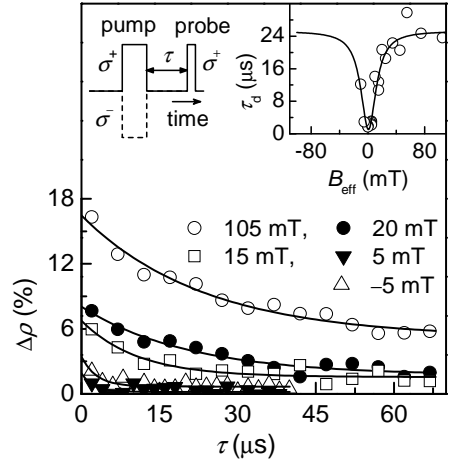}
\caption{Decay of $\Delta \rho$ at a few values of
$B_{\mathrm{ext}}$. Left inset: Schematic of PL pump-probe method.
Right inset: Dependence of $\tau_{\mathrm{d}}$ on
$B_{\mathrm{eff}}$.} \label{fig2}
\end{figure}

Theoretically proposed three-fold decrease of the electron spin
polarization~\cite{merkulovprb65} due to hyperfine interaction
with ``frozen'' NSF was not observed experimentally in case of
very weak optical pumping of the InP QDs. This is related to a
variation of the amplitude and orientation of the NSF in time
(finite lifetime of NSF) which is able to totally destroy the
electron spin polarization if the pump photons rarely come and
restore the polarization. To estimate the lifetime of NSF we
perform a time-domain measurement by using PL pump-probe
technique~\cite{ikezawaprb72,paljpsj75} shown schematically in the
left inset of Fig.~\ref{fig2}. A circularly ($\sigma^{+}$ or
$\sigma^{-}$) polarized strong pump pulse creates spin orientation
of the resident electrons~\cite{paljpsj75}. The electron spin
orientation is then probed by measuring $\Delta \rho$, the
difference of $\rho_{\mathrm{circ}}$ of the probe PL for the
$\sigma^{+}$ and $\sigma^{-}$ pump, as a function of the
pump-probe delay ($\tau$). Figure~\ref{fig2} plots $\Delta \rho$
as a function of $\tau$ for different values of
$B_{\mathrm{ext}}$. The spin decay time $\tau_{\mathrm{d}}$ is
estimated from a fit $\Delta \rho=A \,
\exp(-\tau/\tau_{\mathrm{d}}) + B$ to these data. A plot of
$\tau_{\mathrm{d}}$ as a function of $B_{\mathrm{eff}}$ is shown
in the right inset of Fig.~\ref{fig2}. We consider that near the
minimum where $B_{\mathrm{eff}} \approx 0$, electron spin dynamics
is ruled by the NSF lifetime, which is estimated to be about
1~$\mu$s from our data.

In conclusion, the effect of nuclear spins on electron spin
dynamics in InP QDs is studied. The values of NSF lifetime and of
hyperfine fields $B_{\mathrm{N}}$ and $B_{\mathrm{f}}$ appearing
due to DNP and due to NSF are estimated.

Authors thank V. K. Kalevich and I. Ya. Gerlovin for fruitful
discussions.  The work is partially supported by Grant-in-Aid for
Scientific Research {\#}17$\cdot$5056, {\#}13852003 and
{\#}18204028 from the MEXT of Japan and ``R\&D promotion scheme
funding international joint research'' promoted by NICT of Japan,
by ISTC, grant 2679, by Russian Ministry of Sci{.} {\&} Edu{.},
grant RNP.2.1.1.362 and by RFBR, grant 06-02-17137-a.

\end{document}